\documentclass[prd,showpcs,amsmath,amssymb,nofootinbib,longbibliography,twocolumn,showpacs,notitlepage]{revtex4-1}
\usepackage{multirow}
\usepackage{epsfig}
\usepackage{amsmath}
\usepackage{bm}
\usepackage{times}
\usepackage{graphicx}
\usepackage{color}
\usepackage{slashed}
\usepackage{graphicx}
\usepackage{amsmath} 
\usepackage[latin1]{inputenc}
\usepackage{hyperref}
\usepackage{soul}
\usepackage{multirow}
\usepackage{epsfig}
\usepackage{amsmath}
\usepackage{bm} 
\usepackage{times}
\usepackage{graphicx}
\usepackage{color}
\usepackage{slashed}
\usepackage{graphicx}
\usepackage{amsmath}
\usepackage{tikz}
\usetikzlibrary{positioning,shapes}
\usepackage{relsize} 
\usepackage[latin1]{inputenc}
\usepackage{tikz}
\usetikzlibrary{trees}
\usetikzlibrary{decorations.pathmorphing}
\usetikzlibrary{decorations.markings}
\usetikzlibrary{positioning,arrows}
\usetikzlibrary{decorations.pathmorphing}
\usetikzlibrary{decorations.markings}
\usetikzlibrary{decorations.pathreplacing,calc}
\usetikzlibrary{decorations.pathmorphing,decorations.markings,trees,positioning,arrows}   
\newif\ifmirrorsemicircle
\usetikzlibrary{decorations.pathreplacing,decorations.markings,arrows}

\pgfarrowsdeclarecombine[-5pt]{circled}{circled}{latex}{latex}{o}{o}
\pgfarrowsdeclaredouble{doubled}{doubled}{stealth}{stealth}

\def\bea{\begin{eqnarray}}
\def\eea{\end{eqnarray}}
\def\bean{\begin{eqnarray*}}
\def\eean{\end{eqnarray*}} 
\def\nn{\nonumber}
\def\beaal{\begin{align}}
\def\eeaal{\end{align}}

\begin{document} 
 
\title{Asymmetric Dark Matter from Gravitational Waves}

\author{Bartosz~Fornal}
\affiliation{Department of Chemistry and Physics, Barry University, Miami Shores, Florida 33161, USA\vspace{1mm}}
\author{Erika~Pierre\vspace{1mm}}
\affiliation{Department of Chemistry and Physics, Barry University, Miami Shores, Florida 33161, USA\vspace{1mm}}

\date{\today}

\begin{abstract}
We investigate the prospects for probing asymmetric dark matter models through their gravitational wave signatures. We concentrate on a theory extending the Standard Model gauge symmetry by a non-Abelian group, under which leptons form doublets with new fermionic partners, one of them being a dark matter candidate. The breaking of this new symmetry occurs at a high scale, and results in a strong first order phase transition in the early Universe. The  model  accommodates  baryogenesis in an asymmetric dark matter setting and predicts a 
 gravitational wave signal within the reach of near-future experiments. 
 \end{abstract}

\maketitle

\section{Introduction}
In the past several decades, theoretical and experimental particle physics have brought us incredible insight into how the Universe works at the most fundamental level, with our current knowledge extending down to distances $\sim  10^{-18} \ \rm m$. The Standard Model of elementary particles, formulated in the 1960s \cite{Glashow:1961tr,Higgs:1964pj,Englert:1964et,Weinberg:1967tq,Salam:1968rm} and 1970s \cite{Fritzsch:1973pi,Gross:1973id,Politzer:1973fx}, provides the most comprehensive description of physics at such small scales, with the last piece of the puzzle, the Higgs boson, discovered a decade ago at the Large Hadron Collider  (LHC) \cite{CMS:2012qbp}. Although currently accessible sporadically only in high-energy environments, particle physics effects were not always so elusive. 
Cosmological observations indicate that the Universe is expanding and started off from a state with a large density and temperature, when it was precisely the physics at the small scale that drove its evolution. This introduces additional motivation for exploring particle  physics models, especially in light of the fact that several fundamental questions still remain unanswered. Among the most pressing open issues are the nature of dark matter and the origin of the matter-antimatter asymmetry of the Universe, which both require the existence of new physics, i.e., particles and interactions beyond those described by the
Standard Model.

Observations indicate 
that there is five times more matter in the Universe than what can be attributed to visible matter. The existence of this
 dark matter was inferred from its gravitational interaction with normal matter, first from applying the virial theorem to a galaxy cluster \cite{1933AcHPh...6..110Z,2017arXiv171101693A}, and then  through the measurements of galactic rotation curves \cite{1970ApJ...159..379R}. By now, the evidence for dark matter in the Universe is
overwhelming, with its distribution and abundance precisely determined also from  the cosmic microwave background \cite{Boomerang:2000efg} and gravitational lensing \cite{Gavazzi:2007vw}. Despite those huge advances, 
the mass of the dark matter particle and its non-gravitational interactions with the Standard Model remain a
mystery. It is not even known if the dark matter consists of individual particles or whether it is made up of macroscopic objects like dark quark nuggets \cite{Bai:2018dxf} or primordial black holes \cite{Carr:2009jm,Bird:2016dcv}. 
The allowed  masses for particle dark matter span an extremely  wide range of values, starting from very small ones like in the case of fuzzy dark matter \cite{Press:1989id,Hui:2016ltb}, through intermediate ones like for standard weakly interacting massive particles (WIMPs) \cite{Steigman:1984ac}, up to very large masses for WIMPzillas \cite{Kolb:1998ki,Meissner:2018cay}.
For a review of  particle dark matter candidates, see \cite{Feng:2010gw} and references therein.

In the matter-antimatter asymmetry problem, the question is how at some point  in the early Universe there happened to be slightly more matter than antimatter, despite both being produced  in equal amounts during post-inflationary  reheating. The minimal conditions needed to achieve this include out-of-equilibrium dynamics, as well as violation of  baryon number, charge and the charge-parity symmetry \cite{Sakharov:1967dj}. 
A very attractive class of theories is singled out if one assumes that the ordinary and dark sectors share a common origin. Such a connection is hinted by the fact that the abundances of dark matter 
and ordinary matter are roughly of the same order. This observation 
lies at the heart of theories of asymmetric dark matter \cite{Nussinov:1985xr,Kaplan:1991ah,Hooper:2004dc,Kaplan:2009ag,Petraki:2013wwa,Zurek:2013wia}, in which the 
asymmetries in the dark and visible sectors are generated simultaneously, and  the natural mass scale for
the dark matter particle is on the order of the  proton mass. 
Apart from the GeV-scale dark matter candidate itself, models of asymmetric dark matter contain new heavy particles, which determine the properties of the out-of-equilibrium dynamics. In order to gain access to those heavy states, one would need to construct  higher-energy accelerators, much more powerful than the LHC. However, recently a novel and very promising method of probing such particle physics models emerged.

The  first  detection of gravitational waves  by the Laser Interferometer Gravitational Wave
Observatory within the LIGO/Virgo collaboration \cite{LIGOScientific:2016aoc} was a milestone discovery, made one century after the predictions of general relativity \cite{Einstein:1916vd}, and initiated a renaissance period for gravitational wave  astronomy. 
Although only signals arising from black hole and neutron star mergers have been
observed thus far, a primordial stochastic gravitational wave background carrying information about the
very early period in the evolution of the Universe can also be searched for by the LIGO/Virgo/KAGRA (LVK) detectors. 
Indeed, such a stochastic gravitational wave background could have been produced in several cosmological processes, including first order phase transitions in the early Universe \cite{Kosowsky:1991ua} and inflation \cite{Turner:1996ck}, or through the dynamics of topological defects like cosmic strings \cite{Vachaspati:1984gt,Sakellariadou:1990ne} and domain walls \cite{Hiramatsu:2010yz}. In this study, we focus on gravitational waves from phase transitions.
The LVK detectors' frequency range coincides with the scale of new physics triggering the first order phase transition  $\sim \mathcal{O}(10\!-\!100) \ {\rm PeV}$.
This reach 
will
extend towards lower scales
and have better sensitivity  with future experiments like the 
Laser Interferometer Space Antenna  \cite{Audley:2017drz}, Einstein Telescope \cite{Punturo:2010zz}, DECIGO \cite{Kawamura:2011zz}, Cosmic Explorer \cite{Reitze:2019iox}, and Big Bang Observer \cite{Crowder:2005nr}.

The origin of the stochastic gravitational wave background from first order phase transitions is relatively well-understood. 
The energy density at each point in the Universe is determined by the minimum of the effective
potential, which depends on the details of the particle physics model and the temperature.
At high temperatures, the potential has a global minimum located  at zero field value (false vacuum).
As the Universe cools down, the potential can develop another minimum at a nonzero field value (true
vacuum) with a lower energy density than the false vacuum. If a potential barrier separating the two vacua exists, the Universe is  temporarily trapped in the
false vacuum; however, due to thermal fluctuations or via quantum tunneling, it eventually undergoes a first order phase transition
to the true vacuum. This process corresponds to nucleating bubbles of true
vacuum in different patches of the Universe, which then expand and fill  the 
entire space. For this process to be efficient, the bubble nucleation rate  must be larger than 
 the Hubble expansion. The nucleation
rate is determined by the shape of the effective potential -- it depends on the Euclidean
bounce action for the expanding bubble solution (saddle point configuration) interpolating between the two vacua. Once the nucleation starts, 
gravitational waves are generated from 
bubble wall collisions, sound  waves in the plasma, and
magnetohydrodynamic turbulence.

At the particle physics level, a first order phase transition is triggered by spontaneous symmetry breaking. Many attractive extensions of the Standard Model  exhibit an increased symmetry of the Lagrangian at high energies, which breaks down  to  ${\rm SU}(3)_c \times {\rm SU}(2)_L \times {\rm U}(1)_Y$ at low energies. Given the expected gravitational wave signatures of first order phase transitions, detectors like LVK and future gravitational wave experiments are in a great position  to test such theories.
Indeed, a plethora of  particle physics models experiencing spontaneous symmetry breaking have already been explored in the literature regarding their gravitational wave  signatures from first order phase transitions (see \cite{Caldwell:2022qsj} and references therein), including  theories with new physics at the electroweak scale
\cite{Grojean:2006bp,Vaskonen:2016yiu,Dorsch:2016nrg,Bernon:2017jgv,Baldes:2018nel,Chala:2018ari,Alves:2018jsw,Han:2020ekm,Azatov:2021ifm,Benincasa:2022elt}, neutrino seesaw models  \cite{Brdar:2018num,Okada:2018xdh,DiBari:2021dri,Zhou:2022mlz}, baryon/lepton   number violation \cite{Baldes:2017rcu,Hasegawa:2019amx,Fornal:2020esl}), 
grand unified theories \cite{Croon:2018kqn,Huang:2020bbe,Okada:2020vvb}, 
dark gauge groups \cite{Schwaller:2015tja,Breitbach:2018ddu,Croon:2018erz,Hall:2019ank}, models with conformal invariance \cite{Ellis:2020nnr,Kawana:2022fum},
axions \cite{Dev:2019njv,VonHarling:2019rgb,DelleRose:2019pgi}, supersymmetry \cite{Craig:2020jfv,Fornal:2021ovz},
and new flavor physics \cite{Greljo:2019xan,Fornal:2020ngq}.

Models providing explanations for the matter-antimatter asymmetry of the Universe are particularly good candidates to search  for in gravitational wave experiments, since the out-of-equilibrium dynamics needed to generate the baryon asymmetry  is usually triggered by a first order phase transition, which is precisely one of the processes expected to result in a stochastic  gravitational wave background. Such signatures have been investigated in the case of electroweak baryogenesis models, in which new states appear at the scale of hundreds of GeV \cite{Vaskonen:2016yiu,Dorsch:2016nrg,Baldes:2017rcu}. 
In this paper we focus our investigation on theories of asymmetric dark matter with new states at the PeV scale rather than the electroweak scale. This shifts the expected signal to future sensitivity regions of the Einstein Telescope and Cosmic Explorer.
For concreteness, we perform our analysis based on the theory  introduced in \cite{Fornal:2017owa}, in which baryon number violation proceeds through a new type of instanton interactions arising from the non-Abelian nature of the gauge extension of the Standard Model.
The model not only exhibits a strong first order phase transition, but also predicts the formation of domain walls in the early Universe.

\section{Model}

The model we consider \cite{Fornal:2017owa}  is based on the gauge group 
\bea\label{group}
{\rm SU}(3)_c \times {\rm SU}(2)_L \times {\rm U}(1)_Y \times {\rm SU}(2)_\ell \ .
\eea
The  quarks are singlets under ${\rm SU}(2)_\ell$, whereas the leptons are the upper components of ${\rm SU}(2)_\ell$
doublets. Denoting the new fields, other than the right-handed neutrinos, by tildes and primes, the leptonic fields of the model (for each family) have the following  quantum numbers under the group in Eq.\,(\ref{group}):
\bea
 (l_L  \ \ \  \tilde{l}_L)^T\, \equiv \ \,\hat{l}_L&=& (1,2,-\tfrac12,2) \ , \ \ \ l'_R= (1,2,-\tfrac12,1) \ , \ \ \ \ \nn\\
 (e_R  \ \  \tilde{e}_R)^T \equiv \  \hat{e}_R &=& (1,1,-1,2) \ ,  \ \ \ e'_L= (1,1,-1,1) \ , \nn\\
(\nu_R  \ \  \tilde{\nu}_R)^T\equiv \ \hat{\nu}_R &=& (1,1,0,2) \ , \ \ \ \ \ \ \nu'_L = (1,1,0,1) \ .
\eea

In order to spontaneously break ${\rm SU}(2)_\ell$  and accommodate a successful  mechanism for baryogenesis, two complex ${\rm SU}(2)_\ell$ doublet scalar fields are introduced, $\Phi_1$ and $\Phi_2$. The general form of the scalar potential is given by 
\bea\label{scalar_pot}
&&\hspace{-7mm}V(\Phi_1,\Phi_2) = m_1^2 |\Phi_1|^2 +  m_2^2 |\Phi_2|^2 - (m_{12}^2 \Phi_1^\dagger \Phi_2+{\rm h.c.}) \nn\\
&+& \lambda_1 |\Phi_1|^4 +  \lambda_2 |\Phi_2|^4 + \lambda_3|\Phi_1|^2 |\Phi_2|^2 +  \lambda_4 |\Phi_1^\dagger \Phi_2|^2 \nn\\
&+&  \big[\big(\tilde\lambda_5 |\Phi_1|^2  + \tilde\lambda_6|\Phi_2|^2 +\tilde\lambda_7\Phi_1^\dagger \Phi_2\big)\Phi_1^\dagger \Phi_2+{\rm h.c.}\big] \ , \ \ \ 
\eea
where the parameters $m_{12}^2$, $\tilde\lambda_5$, $\tilde\lambda_6$, and $\tilde\lambda_7$ are complex.
The scalar fields $\Phi_1$ and $\Phi_2$ develop vacuum expectation values (vevs) $v_1$ and $v_2$, respectively, upon which the ${\rm SU}(2)_\ell$  symmetry is broken and one is left with the Standard Model gauge group. Those fields can be written as
\bea
\Phi_j = \begin{pmatrix}
c_{1j}+ i c_{2j} \\
\frac{1}{\sqrt2} (v_j+p_{j} + i a_{j})
\end{pmatrix} 
\eea
for $j=1,2$, where $p_{j}$, $a_{j}$, $c_{1j}$ and $c_{2j}$ are real fields. 
To simplify the notation, we introduce the parameters
\bea
v_\ell \equiv \sqrt{v_1^2+v_2^2} \ , \ \ \ \ \  \tan\beta \equiv \frac{v_2}{v_1} \ .
\eea
In the regime $v_\ell \sim \mathcal{O}(1-1000) \  {\rm PeV}$, the LEP-II experimental bound  $v_\ell \gtrsim 1.7 \ {\rm TeV}$ \cite{Schwaller:2013hqa} is easily satisfied. 
\vspace{1mm}

The Yukawa interactions in the model are given by
\bea\label{Yukawa}
 \mathcal{L}_{\rm Y} &=&  \sum_j\Big( Y_{l}^{ab}\,\bar{\hat{l}}_L^a \, \hat{\Phi}_j\,  {l'}_{\!\!R}^{b}+ Y_{e}^{ab} \, \bar{\hat{e}}_R^a \, \hat{\Phi}_j\, {e'}_{\!\!L}^{b}
   + Y_{\nu}^{ab}\,\bar{\hat{\nu}}_R^a \, \hat{\Phi}_j\, {\nu'}_{\!\!L}^{b}\Big)\nonumber\\  [-3pt]
 &+&   y_e^{ab}\, \bar{\hat{l}}_L^a \,H \,\hat{e}_R^b + y_\nu^{ab}\, \bar{\hat{l}}_L^a\, \tilde{H} \,\hat{\nu}_R^b+ {y'}_{\!\!e}^{ab}\, \bar{l'}_{\!\!R}^a \,H \,{e'}_{\!\!L}^b  \nonumber\\
 &+& {{y'}}_{\!\!\nu}^{ab}\, \bar{l'}_{\!\!R}^a\,\tilde{H} \, {\nu'}_{\!\!L}^b + {\rm h.c.} \ ,
\eea
with an implicit sum over the flavor indices $a,b$. The terms in the first line of Eq.\,({\ref{Yukawa}}), involving the matrices $Y_l$, $Y_e$ and $Y_\nu$, result in vector-like masses for the new fermions. The Yukawa matrices $y_e$ and $y_\nu$ provide masses to the Standard Model charged leptons and neutrinos, whereas $y'_e$ and $y'_\nu$ lead to an additional contribution to the new fermion masses.  Under the phenomenologically natural assumption,
\bea\label{assumption}
Y_{l,e,\nu} v_\ell \gg y_{l,e,\nu} \,v_H \ , \ \ \ \ Y_{l,e,\nu} v_\ell \gg y'_{l,e,\nu} v_H \ , 
\eea
where $v_H$ is the Higgs vev, all constraints from electroweak precision data are satisfied.

After ${\rm SU}(2)_\ell$ breaking, there exist six electrically charged and six neutral new fermionic states $f'$. 
It was demonstrated in \cite{Fornal:2017owa} that a remnant  ${\rm U}(1)_\ell$ symmetry forbids the new fermions from decaying to Standard Model particles. As a result, if the lightest of those states, say $\chi$, is electrically neutral, it becomes a good dark matter candidate. The condition in Eq.\,(\ref{assumption}) assures that the electroweak doublet contribution to $\chi$ is small, and, to a good approximation,
\bea
\chi_L \approx \nu'_L \ , \ \ \ \  \chi_R \approx \tilde\nu_R \ . 
\eea
For simplicity, we assume that the elements of the matrices $Y_l$, $Y_e$ and $Y_\nu$  are small, $(Y_{l,e,\nu})_{ij}\ll 1$. Nevertheless, given the $\sim \mathcal{O}(10)\,\rm PeV$ symmetry breaking scale, the masses of the new fermions can still be large. In particular, one may envision a scenario with  11 heavy new fermions with masses $m_{f'}\sim \mathcal{O}(1)\,{\rm PeV}$ and one light dark matter state $\chi$ with mass $m_\chi \approx 5 \ {\rm GeV}$ (see Sec.\,\ref{BandDM}).

The gauge sector contains three new vector gauge bosons: $Z'$, $W'_1$, and $W'_2$. Denoting by $g_\ell$ the ${\rm SU}(2)_\ell$ gauge coupling, their masses are 
\bea
m_{Z'\!,W'_{1,2}} = \tfrac12 \,g_\ell \, v_\ell \ .
\eea
The new gauge bosons have no direct couplings to quarks, so there do not exist any
 unsuppressed tree-level  diagrams contributing to dark matter direct detection. Although processes relevant for direct detection do arise at the loop level, the resulting 
limits set by the CDMSlite experiment  \cite{SuperCDMS:2015eex} are much weaker than the aforementioned LEP-II constraint.

The scalar content of the theory consists of two real $CP$-even states, one real $CP$-odd state, and two complex conjugated states, which we denote respectively by
\bea
P_1, \ P_2 \ , A \ , C_1 \ , C_2 \ .
\eea
Their masses depend on the parameters of the scalar potential in Eq.\,(\ref{scalar_pot}) and, without  tuning, are naturally at the scale $\sim v_\ell$. As argued in Sec.\,\ref{BandDM}, the $CP$-odd scalar $A$ is chosen to be lighter than  $\chi$, so that there exist efficient dark matter annihilation channels.
The field-dependent masses for all those particles are discussed in Sec.\,\ref{eff_pot}, including  the ones  for the  three Goldstone bosons  $G$, $G_1$, and $G_2$.

\section{Effective potential}
\label{eff_pot}

In order to investigate the dynamics of the phase transition, one needs to determine  the shape of the effective potential. 
In contrast to  \cite{Fornal:2017owa}, in this work we will not assume $v_1 \gg v_2$. 
Given the large number of parameters, to make our analysis more transparent we set $\lambda_3, \lambda_4, \tilde{\lambda}_5, \tilde\lambda_6 = 0 $. The conditions required for vacuum stability reduce then to
\bea
\lambda_1,\lambda_2 >0 \ \ \ \ {\rm and } \ \ \ \  |\tilde\lambda_7 | < \sqrt{\lambda_1\lambda_2} \ . \ \ \ \ \
\eea
With a small nonzero parameter $m_{12}^2$, the theory exhibits  a softly broken $\mathcal{Z}_2$ symmetry defined by the transformation
\bea
\label{transformation}
\Phi_1 \to \Phi_1 \ , \ \ \ \Phi_2 \to - \Phi_2 \ .
\eea

The effective potential is a function  of the classical background fields $(\phi_1,\phi_2)$ and  consists of 3 contributions: tree-level, one-loop Coleman-Weinberg,  and finite temperature,
\bea\label{fulll}
&&\hspace{-5mm}V_{\rm eff}(\phi_1,\phi_2, T) \nn\\
&=& V_{\rm tree}(\phi_1,\phi_2) + V_{\rm loop}(\phi_1,\phi_2) + V_{\rm temp}(\phi_1,\phi_2, T)\ . \ \ \ \ \  \ \ \ 
\eea
The tree-level contribution, upon expressing the parameters $\mu_{1}^2$ and $\mu_{2}^2$ in terms of $v_\ell$, $\beta$, $\mu_{12}^2$ and $\tilde\lambda_7$ using the minimization conditions, takes the form
\bea
&&\hspace{-7mm}V_{\rm tree}(\phi_1,\phi_2)= \tfrac14 \lambda_1 \phi_1^4 +  \tfrac14 \lambda_2 \phi_2^4 - \mu_{12}^2 \phi_1\phi_2 +\tfrac12\lambda_7 \phi_1^2\phi_2^2\nn\\
&+& \tfrac12\big[ \,\mu^2_{12}\tan\!\beta -\lambda_1 v_\ell^2 \cos^2\!\beta -\lambda_7v_\ell^2\sin^2\!\beta \big]\phi_1^2 \ \ \nn\\
  &+& \tfrac12\big[\,\mu^2_{12}\cot\!\beta-\lambda_2v_\ell^2\sin^2\!\beta -\lambda_7v_\ell^2\cos^2\!\beta \big]\phi_2^2 \ ,
\eea
where $\mu_{12}^2 = {\rm Re}(m_{12}^2)$ and $\lambda_7 = {\rm Re}(\tilde\lambda_7)$. 
Adopting the $\overline{\rm MS}$ renormalization scheme, the Coleman-Weinberg term is \cite{Quiros:1999jp}
\bea
  &&\!\!\!\!\!\!\!V_{\rm loop}(\phi_1,\phi_2)  \nn\\
  &=& \sum_i \frac{n_i}{64\pi^2} m_i^4(\phi_1,\phi_2) \left[\log\left(\frac{m_i^2(\phi_1,\phi_2)}{\Lambda^2}\right)-c_i\right] \! , \ \ \ \ \ \ \ \ 
\eea
where the sum includes all particles charged under ${\rm SU}(2)_\ell$, $m_i(\phi_1,\phi_2)$ are their field-dependent masses, $n_i$ denotes their number of degrees of freedom (with a negative sign for fermions), $c_i=3/2$ for fermions and scalars, $c_i=5/6$ for vector bosons, and $\Lambda$ is the renormalization scale.

For  the gauge bosons in the theory, the squared field-dependent masses are
\bea
m_{Z'\!,W'_{1,2}} ^2(\phi_1,\phi_2) = \tfrac14 \,g^2_\ell(\phi_1^2+\phi_2^2)\ .
\eea
Because of our simplifying assumption regarding the structure of the Yukawa matrices, i.e., $(Y_{l,e,\nu})_{ij}\ll 1$, the  field-dependent masses for the fermions are much smaller than those for the gauge bosons, and we will neglect them. 

In the case of scalars, the squared field-dependent masses are given by the eigenvalues of three $2\times 2$ matrices:  $\mathcal{M}_P^2$ for the $CP$-even states $P_1$ and $P_2$, 
\bea
(\mathcal{M}_P^2)_{11} &=& \lambda_1\big(3 \phi_1^2 - v_\ell^2\cos^2\!\beta\big)+ \mu_{12}^2 \tan\!\beta  \nn\\
&+&   \lambda_7\big(\phi_2^2 -  v_\ell^2\sin^2\!\beta\big)\ , \nn\\
(\mathcal{M}_P^2)_{12} &=&(\mathcal{M}_P^2)_{21} = 2\lambda_7 \phi_1\phi_2 - \mu_{12}^2 \ , \nn\\ [2pt]
(\mathcal{M}_P^2)_{22} &=& \lambda_2\big(3 \phi_2^2 - v_\ell^2\sin^2\!\beta\big)+ \mu_{12}^2 \cot\!\beta \ \ \ \ \ \ \ \ \ \ \ \ \ \ \ \  \nn\\
&+&  \lambda_7\big(\phi_1^2 -   v_\ell^2\cos^2\!\beta\big)  \ , 
\eea
$\mathcal{M}_A^2$ for the $CP$-odd state $A$ and the Goldstone boson $G$,
\bea
(\mathcal{M}_A^2)_{11} &=& \lambda_1\big(\phi_1^2 - v_\ell^2\cos^2\!\beta\big)+ \mu_{12}^2 \tan\!\beta  \nn\\
&-&   \lambda_7\big(\phi_2^2 +  v_\ell^2\sin^2\!\beta\big)\ , \nn\\
(\mathcal{M}_A^2)_{12} &=&(\mathcal{M}_A^2)_{21} = 2\lambda_7 \phi_1\phi_2 - \mu_{12}^2 \ , \nn\\ [2pt]
(\mathcal{M}_A^2)_{22} &=& \lambda_2\big(\phi_2^2 - v_\ell^2\sin^2\!\beta\big)+ \mu_{12}^2 \cot\!\beta \ \ \ \ \ \ \ \ \ \ \ \ \ \ \ \  \nn\\
&-&  \lambda_7\big(\phi_1^2 +   v_\ell^2\cos^2\!\beta\big)  \ , 
\eea
and $\mathcal{M}_C^2$ for the complex states $C_{1,2}$ and the Goldstones $G_{1,2}$,
\bea
(\mathcal{M}_C^2)_{11} &=& \lambda_1\big(\phi_1^2 - v_\ell^2\cos^2\!\beta\big)+ \mu_{12}^2 \tan\!\beta  -   \lambda_7 \,  v_\ell^2\sin^2\!\beta\ , \nn\\
(\mathcal{M}_C^2)_{12} &=&(\mathcal{M}_C^2)_{21} = 2\lambda_7 \phi_1\phi_2 - \mu_{12}^2 \ , \nn\\ [2pt]
(\mathcal{M}_C^2)_{22} &=& \lambda_2\big(\phi_2^2 - v_\ell^2\sin^2\!\beta\big)+ \mu_{12}^2 \cot\!\beta - \lambda_7\,  v_\ell^2\cos^2\!\beta  \ .\nn\\
\eea

The finite temperature contribution to the potential is \cite{Quiros:1999jp}
\bea
&&V_{\rm temp}(\phi_1,\phi_2, T) \nn\\[2pt]
&&= \frac{T^4}{2\pi^2} \sum_i n_i \int_0^\infty dy \,y^2 \log\left(1\mp e^{-\sqrt{m_i^2(\phi_1,\phi_2)/T^2 + y^2}}\right) \ \ \ \ \ \nn\\
&& +\, \frac{T}{12\pi} \sum_j n'_j \Big\{m_j^3(\phi_1,\phi_2) - [m^2(\phi_1,\phi_2) + \Pi(T)]_j^{\frac32}\Big\} \, , \ \ \ \ \ \ 
\eea
where in the second line the minus sign corresponds to bosons and the plus sign to fermions, the sum over $i$ incorporates all particles with field-dependent masses, the sum over $j$ includes only bosons, $n_i$ denotes the number of degrees of freedom for a given particle, $n'_j$ is the number of all degrees of freedom in the case of  scalars and solely  longitudinal ones for  vector bosons, $\Pi(T)$ is the thermal mass matrix, and $[m^2(\phi_1,\phi_2) + \Pi(T)]_j$ are the eigenvalues of the matrix $[m^2(\phi_1,\phi_2) + \Pi(T)]$.
The $2\times2$ thermal mass matrix $\Pi(T)$ is diagonal and identical for the four pairs of scalars $(P_1, P_2)$, $(A, G)$, $(C_1,G_1)$, $(C_2,G_2)$. It is given by
\bea
\Pi(T)= 
\begin{pmatrix}
\frac3{16} g_\ell^2 +\tfrac12\lambda_1 & 0  \\
0 & \tfrac3{16} g_\ell^2 + \tfrac12\lambda_2
\end{pmatrix} T^2 \ . \ \ \ \ \ \ \ \ 
\eea
In the case of vector gauge bosons, the thermal masses are
\bea
\Pi_{Z'}(T) = \Pi_{W'_{1,2}}(T) = 2 g_\ell^2T^2 \ .
\eea 

\begin{figure}[t!]
\includegraphics[trim={1cm 0.8cm 1cm 0cm},clip,width=9cm]{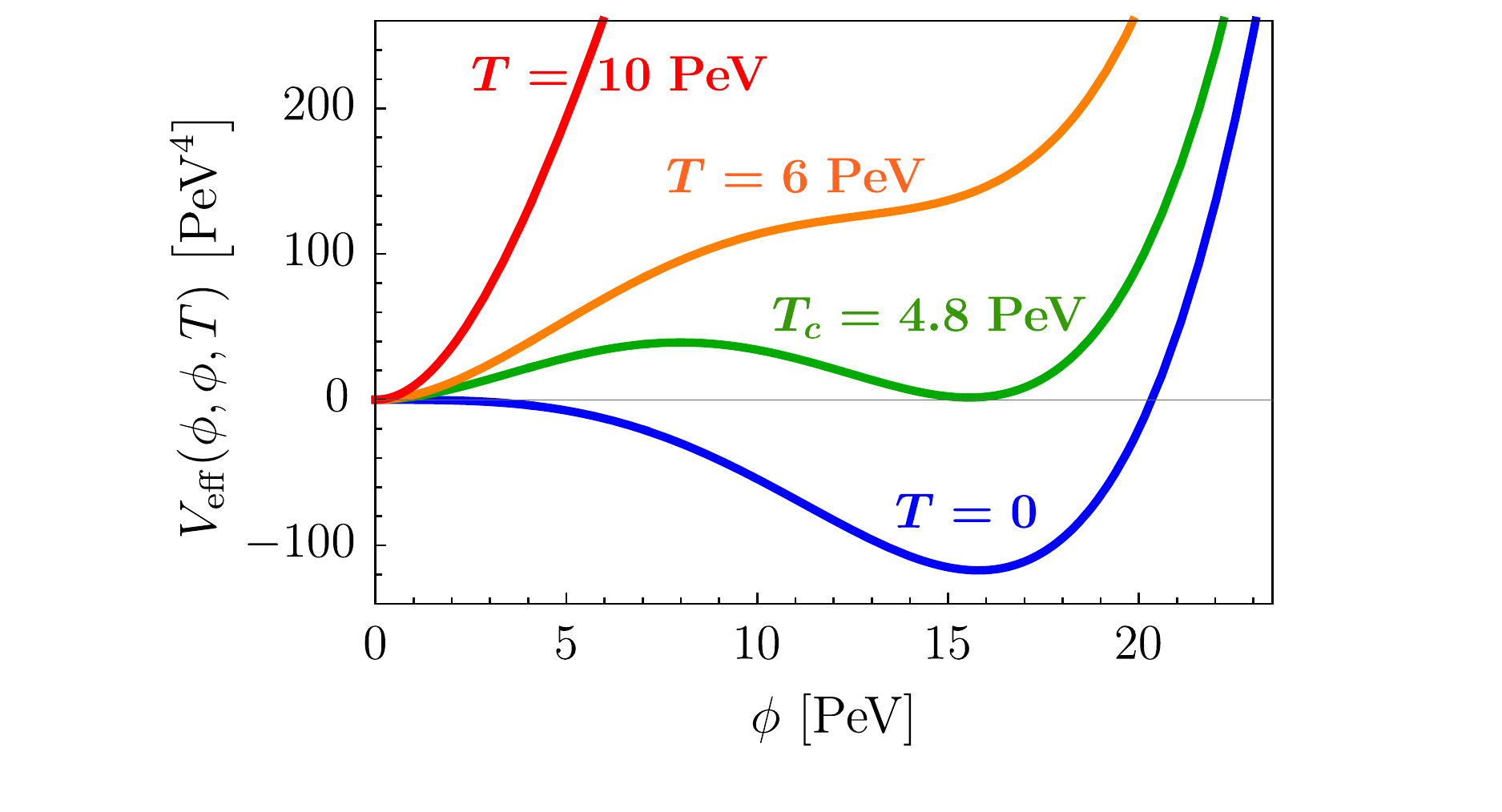} \vspace{-4mm}
\caption{Effective potential $V_{\rm eff}(\phi_1, \phi_2, T)$ along the field direction $\phi_1=\phi_2\equiv \phi$ and normalized to zero at the origin, for the parameter choice discussed in the text and several   values of temperature.}\label{V12}
\end{figure}

As an illustration, Fig.\,\ref{V12} presents a slice of the effective potential along the field direction $\phi_1=\phi_2$ for several different temperatures and for the parameter choice: $\lambda_1=\lambda_2 = 10^{-3}$,  $\beta = \pi/4$, \,$v_\ell = \Lambda = 10 \ \rm PeV$, \,$g_\ell=1$, and small  $|\mu_{12}^2|$, $|\lambda_7|$.

As the temperature decreases, a new local minimum of the effective potential develops away from the origin. Below  the critical temperature $T_c$, this minimum, which we denote by $(\phi_1,\phi_2)_{\rm true1}$, becomes the new  true vacuum of the theory. When the temperature drops further to the so-called nucleation temperature $T_*$,  patches of the Universe undergo a phase transition to this preferred true vacuum. 
Since the new vacuum is separated  by a potential bump from the false vacuum at the origin, the phase transition is first order. Details of the resulting gravitational wave  signal are discussed in Sec.\,\ref{GW_tr}.

The vacuum $(\phi_1,\phi_2)_{\rm true1}$ is not the only minimum with energy density lower than that of  the false vacuum at  the origin.  The effective potential develops four minima which come in two pairs -- the vacua within each pair are related via the approximate $\mathcal{Z}_2$ symmetry of the potential defined in Eq.\,(\ref{transformation}), while the two pairs are related to each other via a gauge symmetry. In particular, they are related through a rephasing transformation of the Lagrangian  fields $\Phi_i \to e^{i\theta}\Phi_i$ ($i=1,2$), making them physically equivalent \cite{Ginzburg:2004vp}. For a detailed discussion of the topology of the scalar potential in two-Higgs doublet models, see \cite{Battye:2011jj}.

As a result, there are only two physically distinct true vacua of the theory, $(\phi_1,\phi_2)_{\rm true1}$ and $(\phi_1, - \phi_2)_{\rm true2}$. Their energy densities differ solely  because of the nonzero $\mathcal{Z}_2$ symmetry breaking terms in the effective potential, i.e., in our case  the term involving the parameter $\mu^2_{12}$. If the energy density difference between those two vacua is large,  the Universe transitions to the vacuum with lower energy density. However, if the splitting is small, i.e., $|\mu_{12}^2| \ll \lambda_{1,2} v_\ell^2$, then a given patch of the Universe can transition to either $(\phi_1,\phi_2)_{\rm true1}$ or $(\phi_1, - \phi_2)_{\rm true2}$\,, leading to the formation of domain walls \cite{Saikawa:2017hiv}. Their subsequent annihilation constitutes another possible source of gravitational radiation.

\section{Baryogenesis and dark matter}
\label{BandDM}

A first order phase transition provides exactly the out-of-equilibrium dynamics needed to generate a matter-antimatter asymmetry of the Universe. 
The remaining requirements, i.e., violation of baryon number, charge, and the charge-parity symmetry, are also present in  the model. As shown in   \cite{Fornal:2017owa}, this  leads to a successful mechanism for baryogenesis, which combines the features of asymmetric dark matter 
\cite{Nussinov:1985xr,Kaplan:1991ah,Hooper:2004dc,Kaplan:2009ag,Petraki:2013wwa,Zurek:2013wia}, Dirac leptogenesis  \cite{Dick:1999je,Murayama:2002je}, and baryon asymmetry generation from an earlier phase transition \cite{Shu:2006mm} (see also \cite{Blennow:2010qp}). 
In this section, we summarize the most important aspects of this proposal.

Baryon number violation in the model is a result of a lepton number asymmetry produced by the nonperturbative dynamics of ${\rm SU}(2)_\ell$ instantons,
which remain active outside the expanding bubble of true vacuum, but are exponentially suppressed inside the bubble.  
As derived in \cite{Fornal:2017owa} (following a similar calculation in \cite{Morrissey:2005uza}), the ${\rm SU}(2)_\ell$ instantons induce the dimension-six interactions
\bea\label{6int}
\mathcal{O}_6 \sim \epsilon_{ij}  &&\hspace{-1.5mm}\Big[(l_L^i \cdot \bar{{\nu}}_R)(l_L^j \cdot \bar{{e}}_R) - (l_L^i \cdot \bar{{\nu}}_R)(\tilde{l}_L^j \cdot \bar{\tilde{e}}_R) \nn\\
&\hspace{-2.5mm}+& (l_L^i \cdot \tilde{l}_L^j)(\bar{\nu}_R \cdot \bar{\tilde{e}}_R)- (l_L^i \cdot \tilde{l}_L^j)(\bar{\tilde{\nu}}_R \cdot \bar{{e}}_R) \nn\\
& \hspace{-2.5mm}+&  (\tilde{l}_L^i \cdot \bar{\tilde{\nu}}_R)(\tilde{l}_L^j \cdot \bar{\tilde{e}}_R) - (\tilde{l}_L^i \cdot \bar{\tilde{\nu}}_R)({l}_L^j \cdot \bar{{e}}_R)\Big], \ \ \ \ \ 
\eea
written for simplicity for a single generation of matter, and with the dot denoting Lorentz contraction. 
Lepton number asymmetry is generated, e.g., via the last term, which gives rise  to the process $\nu_L \tilde{e}_L \to \tilde{\nu}_R e_R$ and results in a violation of lepton number by $\Delta L = -1$. At the same time, due to an existing global ${\rm U}(1)_\chi$ symmetry (see \cite{Fornal:2017owa} for details), this process also leads to the violation of the dark matter number by $\Delta \chi = 1$. 
With a sufficient  amount of $CP$ violation in the model, part of the instanton-generated lepton asymmetry outside the expanding bubble becomes trapped inside the bubble, with a similar process taking place in the dark matter sector.
Quantitatively, the production of the two asymmetries is governed by the diffusion equations \cite{Joyce:1994zn,Cohen:1994ss},
\bea
\dot{\rho}_i = D_i \nabla^2 \rho_i - \sum_j \Gamma_{ij} \frac{\rho_j}{n_j} + \gamma_i \ \ ,
\eea
where $\rho_i$ is the number density for a given type of particles, $D_i$ is the diffusion constant, $\Gamma_{ij}$ is the rate of diffusion, $n_j$ is the number of degrees of freedom (with a minus sign for fermions), and $\gamma_i$ are the $CP$-violating sources. Given our assumption of small new Yukawa couplings $Y\ll 1$, the sources take the form \cite{Riotto:1995hh}
\bea
\gamma_i \approx \frac{\lambda_7 \mu_{12}^2}{32\pi}\frac{\Gamma_{\phi_i} T_*}{m_{\phi_i}^3\!(T_*)} \partial_z \phi_i \ ,
\eea
where $\Gamma_{\phi_i}$ is the decay rate  of $\phi_i$ and the derivative $\partial_z$ is taken along the direction perpendicular to the bubble wall. The strength of the sources determines the amount of lepton and dark matter asymmetries generated. 

In the model under consideration, there are twelve diffusion equations and eight constraints arising from Yukawa and instanton interactions (see \cite{Fornal:2017owa} for details). Given the form of those interactions in Eq.\,(\ref{6int}),  the ratio of the generated lepton and dark matter asymmetries is
\bea\label{LL}
\left| \frac{\Delta L}{\Delta \chi}\right| = 3 \ .
\eea
Upon  the completion of ${\rm SU}(2)_\ell$ breaking, the resulting dark matter asymmetry remains unaltered, but the lepton  asymmetry is partially converted into a baryon asymmetry via the Standard Model electroweak sphalerons \cite{Harvey:1990qw}, which leads to
\bea\label{BB}
\Delta B = \frac{28}{79}\Delta L \ .
\eea

To determine the parameters for which a sufficiently large baryon asymmetry is generated, we solve the diffusion equations for various $\gamma_i$. For consistency with the discussion in Sec.\,\ref{GW_tr}, we adopt the bubble wall velocity equal to the speed of light $(v_w =  c)$, the effective vev $v_\ell =  10 \ {\rm PeV}$, the quartic couplings  $\lambda_i \sim 10^{-4}$, and the temperature $T_* \sim 1 \ {\rm PeV}$.
We find that the observed baryon-to-photon  ratio of \cite{Workman:2022ynf}
\bea
\frac{n_B}{n_\gamma} \approx 6 \times 10^{-10} 
\eea
is obtained when the parameters of the model satisfy 
\bea\label{newnew}
|\lambda_7 \mu_{12}^2|\,Y^2 \sim 10^{-12} \  {\rm PeV^2} \ .
\eea
For example, the following choice of parameters: $\lambda_7 \sim 10^{-6}$, $\mu_{12}^2 \sim 10^{-4} \, {\rm PeV^2}$ and $Y \sim 0.1$, is consistent with our assumptions and leads to the observed matter-antimatter asymmetry of the Universe.

\begin{figure}[t!]
\includegraphics[trim={0cm 0cm 0cm 0},clip,width=8.5cm]{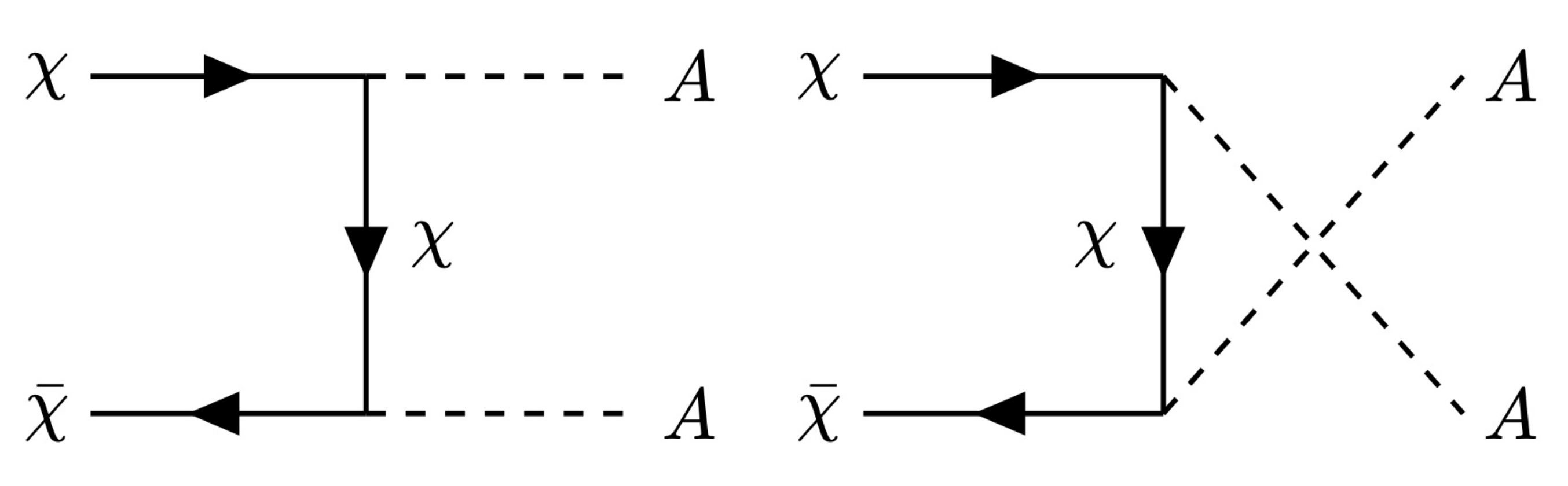} \vspace{-1mm}
\caption{Dark matter annihilation channels.}\label{annihilation}
\end{figure}

Equations (\ref{LL}) and (\ref{BB}) imply that the baryon and dark matter asymmetries are approximately equal at present times. This fixes the dark matter mass to be
\bea
 m_\chi \approx m_p \frac{\Omega_{\rm DM}}{\Omega_b}\bigg|\frac{\Delta B}{\Delta \chi}\bigg| \approx 5 \ {\rm GeV} \ ,
\eea
assuming that it is relativistic at the decoupling temperature. Such a low  mass introduces  the usual challenge for asymmetric dark matter models to annihilate away the symmetric component. The standard solution  is to tune one of the scalars to be light, so that an efficient annihilation channel opens up. This is implemented in the model by arranging for the mass of the $CP$-odd scalar $A$ to be below 5 GeV, which is experimentally allowed \cite{Krnjaic:2015mbs}. This is achieved by choosing a small value of $\lambda_1$, which is also needed for the phase transition to be first order. The resulting annihilation channels for the symmetric component of $\chi$ are shown in Fig.\,{\ref{annihilation}}.

\section{Gravitational waves from\\ \ \ phase transitions}
\label{GW_tr}

As discussed in Sec.\,\ref{eff_pot}, when the temperature becomes sufficiently low, patches of the Universe start undergoing a first order phase transition from the false vacuum at the origin to either of the true vacua:    $(\phi_1,\phi_2)_{\rm true1}$ or $(\phi_1, - \phi_2)_{\rm true2}$. When the $\mathcal{Z}_2$ breaking parameters are small, the expected gravitational wave  signal from a transition to any of those  two vacua is similar. For concreteness, in the subsequent analysis we focus on the transition to $(\phi_1,\phi_2)_{\rm true1}$.

During such a first order phase transition, bubbles of true vacuum are nucleated and gravitational waves are generated through bubble wall collisions, sound shock waves  in the plasma, and magnetohydrodynamic turbulence. 
The phase transition starts when the bubble nucleation rate becomes comparable to the Hubble expansion rate, i.e.,  when $\Gamma(T_*) \sim H^4(T_*)$. The temperature at which  this happens is called the nucleation temperature $T_*$.
The rate for bubble nucleation can be calculated as  \cite{LINDE1983421}
\bea\label{GammaPT}
\Gamma(T) \approx \left(\frac{S_E(T)}{2\pi T}\right)^{\frac32}T^4 \exp\!\left({-\frac{S_E(T)}{T}}\right) \ ,
\eea
where $S_E(T)$ is  the Euclidean action dependent on the shape of the effective potential. Denoting $\vec{\phi} = (\phi_1,\phi_2)^T$, in the case of thermal tunneling $S_E(T)$ is given by the integral
\bea
S_E(T)= \int d^3 x \left[\frac12\left(\partial_\mu \vec{\phi}\right)^2+V_{\rm eff}(\vec\phi, T)\right] , 
\eea
in which $\vec\phi$, assuming spherical symmetry, satisfies the bubble equation of motion,
\bea
\frac{d^2 \vec\phi}{dr^2}+\frac{2}{r}\frac{d\vec\phi}{dr}-\vec\nabla V_{\rm eff}(\vec\phi,T) = 0 \ ,
\eea
with the boundary conditions 
\bea
\frac{d\vec\phi}{dr}\bigg|_{r=0} = 0 \ , \ \ \ \ \ \vec\phi(\infty) = \vec\phi_{\rm false} \ .
\eea
Using Eq.\,(\ref{GammaPT}), the condition for the onset of a phase transition  can be written explicitly as
\bea\label{ST44}
\frac{S_E(T_*)}{T_*}   \approx  4\log\left(\frac{M_{\rm P}}{T_*}\right)  - \log\left[\left(\frac{4\pi^3g_*}{45}\right)^{\!\!2}\!\left(\frac{2\pi \,T_*}{S_E(T_*)}\right)^{\!\!\frac32}\right],\nn\\
\eea
where $M_{\rm P} = 1.22\times 10^{19} \ {\rm GeV}$ is the Planck mass and $g_*$ is the number of degrees of freedom at the temperature $T_*$. 
Equation (\ref{ST44}) serves as the source for determining $T_*$ for a given set of  parameters in the effective potential.

The expected gravitational wave spectrum is fully described by  four quantities:  bubble wall velocity, nucleation temperature, strength of the phase transition, and its duration. Out of those parameters, only the bubble wall velocity is independent of the shape of the effective potential and we  set it to the speed of light, i.e., $v_w = c$. Detailed discussions of how to model $v_w$ more precisely are provided in \cite{Espinosa:2010hh,Caprini:2015zlo}.

The strength of the phase transition is given by the ratio of the energy density of the false vacuum (with respect to the true vacuum) and the energy density of radiation at nucleation temperature,
\bea\label{alpha}
\alpha = \frac{\rho_{\rm vac}(T_*)}{\rho_{\rm rad}(T_*)} \ ,
\eea
where
\bea
\rho_{\rm vac}(T) &=& V_{\rm eff}(\vec\phi_{\rm false},T) -  V_{\rm eff}(\vec\phi_{\rm true},T)\nn\\
&-&T \frac{\partial}{\partial T} {\left[ V_{\rm eff}(\vec\phi_{\rm false},T) -  V_{\rm eff}(\vec\phi_{\rm true},T)\right]} \ \ \ \ \ \ 
\eea
and
\bea
\rho_{\rm rad}(T) = \frac{\pi^2}{30} g_* T^4 \ .
\eea
The inverse of the duration of the phase transition $\tilde\beta$ is
\bea\label{beta}
\tilde{\beta} = T_* \frac{d}{dT} \!\left[\frac{S_E(T)}{T}\right]\bigg|_{T=T_*} \ .
\eea
Numerical simulations have been used to derive empirical formulas describing how  the expected gravitational wave
spectrum from bubble collisions, sound waves, and turbulence depends on the four parameters $v_w$, $T_*$, $\alpha$, and $\tilde\beta$. 
\vspace{1mm}

The contribution from sound waves is given by  \cite{Hindmarsh:2013xza,Caprini:2015zlo}
\bea\label{sw}
h^2 \Omega_{s}(f) \,&\approx&\, \frac{1.9\times 10^{-5}}{\tilde\beta}\frac{(f/f_s)^3}{\big[1+0.75 (f/f_s)^2\big]^{7/2}}\left(\frac{g_*}{100}\right)^{-\frac13}\nn\\
&\times&\left[\frac{\alpha^2}{(1+\alpha)(0.73+0.083\sqrt\alpha + \alpha)}\right]^2 \Upsilon \ ,
\eea
where the formula for the fraction of the latent heat transformed into the plasma's bulk motion derived in  \cite{Espinosa:2010hh} was used, the peak frequency is
\bea\label{swf}
f_s &=& (0.19 \ {\rm Hz} ) \left(\frac{T_*}{1 \ {\rm PeV}}\right)\left(\frac{g_*}{100}\right)^\frac16  \tilde\beta \ ,\ \ \ \ \ \ \ 
\eea
and $\Upsilon$ is  the suppression factor \cite{Ellis:2020awk} for which we adopt the most recent estimate \cite{Guo:2020grp},
\bea
\Upsilon = 1- \frac1{\sqrt{1+\frac{8\pi^{1/3}}{\sqrt3\,\tilde{\beta}}\Big(\frac{\sqrt{(1+\alpha)(0.73+0.083\sqrt\alpha + \alpha)}}{\alpha}\Big)}} \ . \ \ \ \ \ 
\eea

The contribution  to the gravitational wave spectrum from bubble wall collisions can be written as  \cite{Kosowsky:1991ua,Huber:2008hg,Caprini:2015zlo}
\bea\label{col}
h^2 \Omega_{c}(f) \,&\approx&\, \frac{2.5\times 10^{-6}}{\tilde\beta^2}\frac{(f/f_c)^{2.8}}{1+2.8 (f/f_c)^{3.8}}\left(\frac{g_*}{100}\right)^{-\frac13}\nn\\
&\times&\left[\frac{\alpha^2 +0.25\,\alpha\sqrt{\alpha}}{(1+\alpha)(1+0.72\,\alpha)}\right]^2, \ \ \ \ \ \ 
\eea
where the fraction of the latent heat deposited into the bubble front was adopted from \cite{Kamionkowski:1993fg}, and the peak frequency is
\bea
f_c &=& (0.037 \ {\rm Hz} ) \left(\frac{T_*}{1 \ {\rm PeV}}\right)\left(\frac{g_{*}}{100}\right)^\frac16\tilde\beta \ . \ \ \ \ \ \  
\eea

Although turbulence provides a subleading contribution to the signal in the peak region, for completeness we provide the corresponding formula  \cite{Caprini:2006jb,Caprini:2009yp},
\bea\label{tur}
h^2 \Omega_{t}(f) \,&\approx&\, \frac{3.4\times 10^{-4}}{\tilde\beta}\frac{\epsilon^2\ ({f}/{f_t})^{3}}{\big(1+{8\pi f}/{f_*}\big)\big(1+{f}/{f_t}\big)^{{11}/{3}}}\nn\\
&\hspace{-20mm}\times&\hspace{-10mm}\left(\frac{g_*}{100}\right)^{-\frac13}\left[\frac{\alpha^2}{(1+\alpha)(0.73+0.083\sqrt\alpha + \alpha)}\right]^{3/2}\!\!\!\!\!\!,
\eea
again assuming the fraction of the latent heat transformed into the plasma's bulk motion from \cite{Espinosa:2010hh}. 
In the above formula the parameter $\epsilon=0.05$ \cite{Caprini:2015zlo}, the peak frequency
\bea
f_t &=& (0.27 \ {\rm Hz} )\, \frac{\tilde\beta}{v_w}\,\left(\frac{g_*}{100}\right)^\frac16\left(\frac{T_*}{1 \ {\rm PeV}}\right) \ ,
\eea
and the parameter $f_*$ \cite{Caprini:2015zlo},
\bea\label{lasteq}
f_* = (0.17 \ {\rm Hz})\left(\frac{g_*}{100}\right)^\frac16\left(\frac{T_*}{1 \ {\rm PeV}}\right) \ .
\eea

\begin{figure}[t!]
\includegraphics[trim={2.2cm 0.5cm 2cm 0cm},clip,width=9cm]{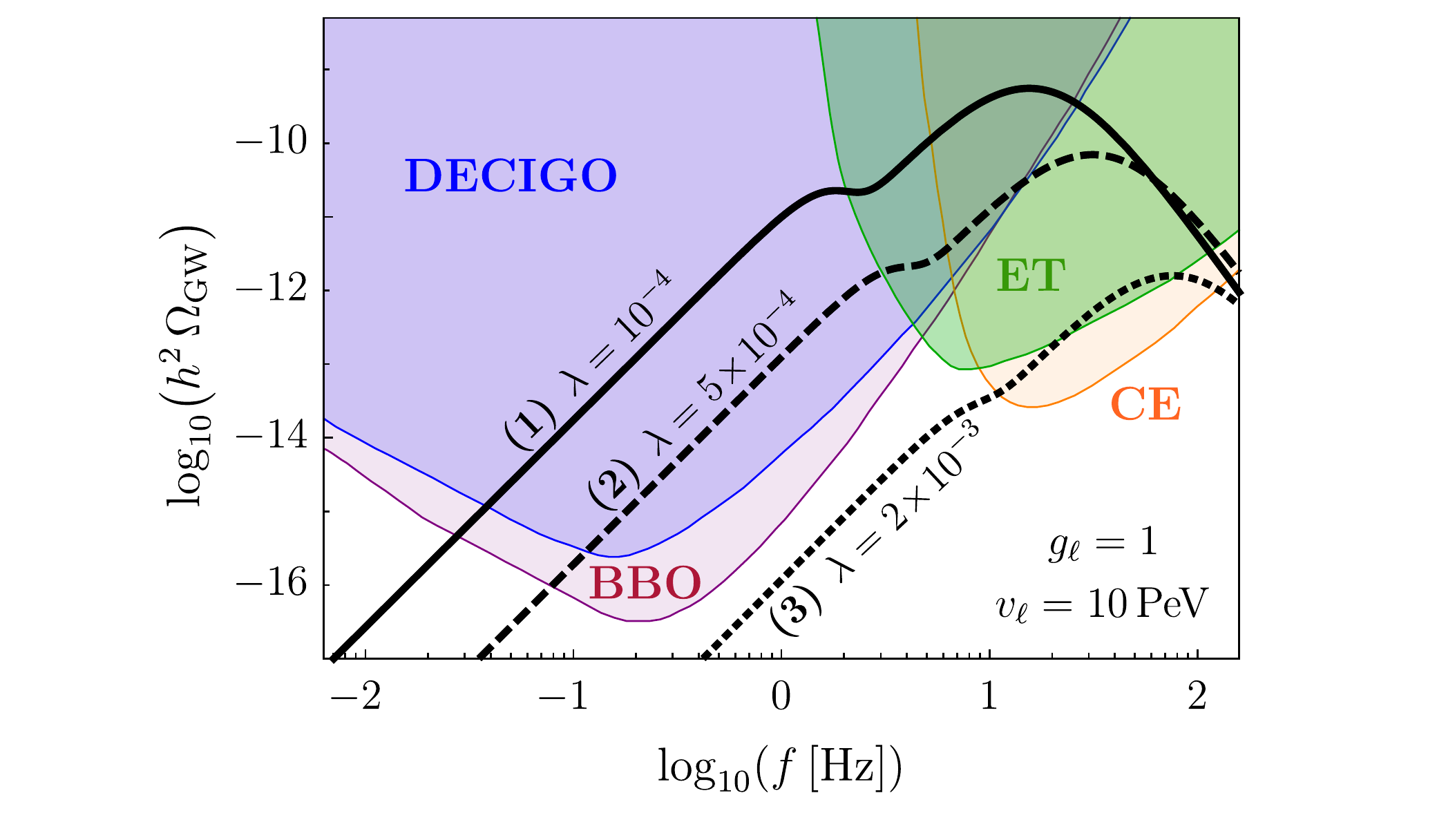} \vspace{-5mm}
\caption{Gravitational wave signatures of  the ${\rm SU}(2)_\ell$ model for the vev $v_\ell = 10 \, \rm PeV$, gauge coupling $g_\ell = 1$, and several values  of the quartic coupling: $\lambda = \{10^{-4}$, $5\!\times\!10^{-4}$, $2\!\times\! 10^{-3}\}$, as described in the figure. Sensitivities of future gravitational wave detectors are also shown:  Einstein Telescope \cite{Sathyaprakash:2012jk} (green), 
Cosmic Explorer \cite{Reitze:2019iox} (orange),  DECIGO \cite{Yagi:2011wg} (blue), and Big Bang Observer \cite{Yagi:2011wg} (purple). The  phase transition parameters corresponding to the curves $(1)$, $(2)$ and $(3)$ are provided in Table \ref{tableOK}.}\label{PT}\vspace{3mm}
\end{figure}

\begin{table}[h!] 
\begin{center}
\begingroup
\setlength{\tabcolsep}{6pt} 
\renewcommand{\arraystretch}{1.5} 
\begin{tabular}{ |c| |c |c |c|| c| c|c|} 
\hline
& \multicolumn{3}{|c||}{Lagrangian parameters}  & \multicolumn{3}{|c|}{Signal parameters} \\ 
\hline
\hline
\!Curve\! & $v_\ell$ & $\lambda$ &   $g_\ell$  & $\alpha$ & $\tilde\beta$   & $T_*$ \\ 
\hline
\hline
$(1)$ & $10 \ {\rm PeV}$   &      $10^{-4}$  &  $1.0$ & \,2.0\, & 70 & $1.2  \ {\rm PeV}$ \\[1pt]
\hline
$(2)$ & $10 \ {\rm PeV}$   &    $5\times 10^{-4}\!$  & $1.0$ &    0.8 & 110 & $1.5  \ {\rm PeV}$ \\[1pt]
\hline
$(3)$ &$10 \ {\rm PeV}$   &     $2 \times 10^{-3}\!$  &  $1.0$ & 0.2  &  \ 200 \ & $ 2.0 \ {\rm PeV}$ \\[1pt]
\hline
\end{tabular}
\endgroup
\end{center}
\vspace{-4mm}
\caption{Values of the phase transition parameters  $\alpha$, $\tilde\beta$ and $T_*$ for the three benchmark gravitational wave signatures shown  in Fig.\,\ref{PT}.}\vspace{1mm}
\label{tableOK}
\end{table}

To determine the  gravitational wave spectra of the ${\rm SU}(2)_\ell$ model, we used the software {\fontfamily{cmtt}\selectfont
anybubble} \cite{Masoumi:2016wot} to compute  the Euclidean action $S_E$ as a function of temperature for various  parameter choices in the  effective potential given by Eq.\,(\ref{fulll}). For simplicity, in our analysis we set the  quartic couplings to be equal, $\lambda_1 = \lambda_2\equiv \lambda$, and we assumed the same for the vevs, $v_1 = v_2$. As mentioned earlier, we took $|\mu_{12}^2|$ and $|\lambda_7|$ to be small. Under those assumptions,  the effective potential is fully described just by the four parameters $(v_\ell, \lambda, g_\ell, T)$. We then numerically determined the nucleation temperature $T_*$ for each case via Eq.\,(\ref{ST44}), and calculated the parameters $\alpha$ and $\tilde\beta$ using Eqs.\,(\ref{alpha}) and (\ref{beta}), to finally arrive at the expected gravitational wave signal,
\bea
h^2\Omega_{\rm GW}(f) = h^2\Omega_{s}(f) +h^2\Omega_{c}(f) +h^2\Omega_{t}(f) \ ,
\eea 
using the expressions in Eqs.\,(\ref{sw})--(\ref{lasteq}).

The resulting gravitational wave signatures, for the three representative sets of parameters listed in Table \ref{tableOK}, are shown in Fig.\,\ref{PT}. In all cases the leading contribution around the peak region comes from sound waves and is given by Eq.\,(\ref{sw}). The smaller bump towards lower frequencies reflects the bubble collision contribution from Eq.\,(\ref{col}). The position of the peak of each signal is proportional to the nucleation temperature, thus signatures corresponding to phase transitions happening at energies higher than $10 \ {\rm PeV}$ would be shifted towards higher frequencies. The peak frequency also has a linear dependence on the parameter $\tilde\beta$. The height of the signal peak is determined by both $\alpha$ and $\tilde\beta$: for larger $\alpha$ the  signal is stronger, whereas for larger $\tilde\beta$ the signal is weaker.

\begin{figure}[t!]
\includegraphics[trim={1cm 0.8cm 2.0cm 2.3cm},clip,width=8cm]{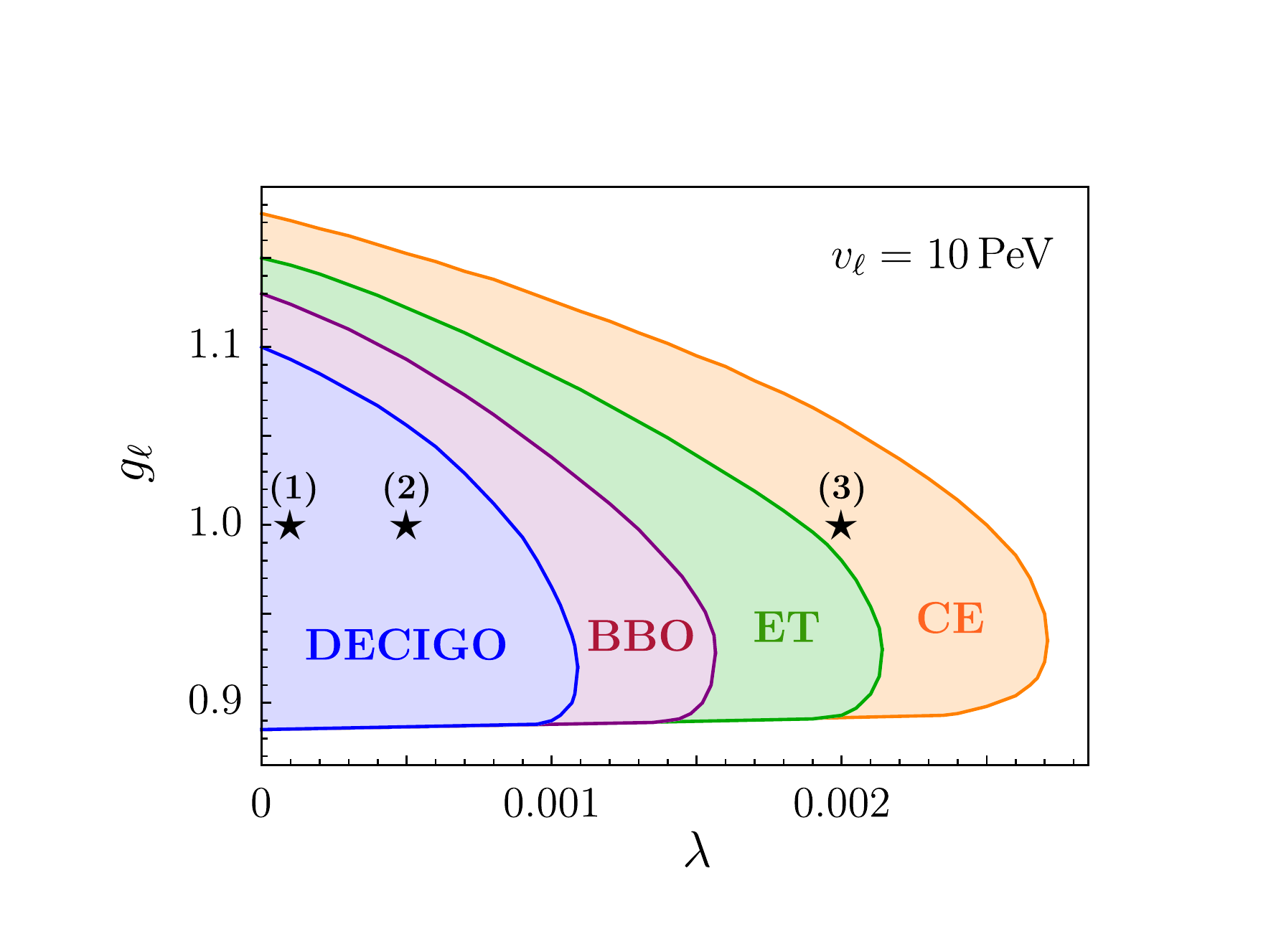} \vspace{-2mm}
\caption{Regions of parameter space $(\lambda, g_\ell)$ of the ${\rm SU}(2)_\ell$ model for $v_\ell = 10 \, {\rm PeV}$ corresponding to a signal detectable, upon one year of data collecting, with a signal-to-noise ratio of at least five, in the experiments:  DECIGO  (blue), Big Bang Observer (blue and purple),  Einstein Telescope (blue, purple and green), Cosmic Explorer (entire shaded region). The  stars denote the benchmark parameters in Fig.\,\ref{PT}.\vspace{-2mm}}\label{scan}
\end{figure}

Depending on the parameter values, the signal of the model with a symmetry breaking scale $v_\ell \sim {\mathcal{O}} (1\!-\!1000) \, {\rm PeV}$ can fall within the sensitivity range of four planned gravitational wave experiments:  Einstein Telescope, Cosmic Explorer,\break DECIGO and Big Bang Observer. The largest signal strength corresponds to a small quartic coupling $\lambda$. In this limit the tree-level term in the effective potential becomes small, and the shape of $V_{\rm eff}$ is determined by the one-loop Coleman-Weinberg term and finite temperature effects. This is known as the supercooling regime \cite{DelleRose:2019pgi,Ellis:2020nnr,Kawana:2022fum}, characterized by a small $\tilde\beta$ and large $\alpha$, which leads to an enhanced gravitational wave signal. For some particle physics models, this scenario can already be searched for in the existing  LVK data \cite{LIGO_FOPT}. 
 
To assess how likely it is for phase transitions in the model to produce a detectable gravitational wave signal,  we performed a scan over the parameters $(\lambda, g_\ell)$ for the vev  fixed at $v_\ell = 10 \ {\rm PeV}$, and  determined the regions corresponding to a signal-to-noise ratio of at least five after a single year of data collecting with the  Einstein Telescope, Cosmic Explorer, DECIGO and Big Bang Observer. The results of the scan are shown in Fig.\,\ref{scan}. A large portion of the ${\rm SU}(2)_\ell$ model parameter space leading to a first order phase transition will be probed by those experiments, with DECIGO and Big Bang Observer being able to probe also lower symmetry breaking scales.

Finally,  as discussed in Sec.\,\ref{eff_pot}, the model predicts also a gravitational wave signal  from domain walls. Indeed, following the analysis of Sec.\,\ref{BandDM}, a successful mechanism for baryogenesis favors a small $\mathcal{Z}_2$ breaking parameter $\mu_{12}^2$, which leads to a near-degeneracy between the  vacua $(\phi_1,\phi_2)_{\rm true1}$ and $(\phi_1, - \phi_2)_{\rm true2}$ , resulting in the production of domain walls in the early Universe. Their  subsequent annihilation gives rise to a gravitational wave background of a predictable  shape \cite{Saikawa:2017hiv}. 
Nevertheless, given the relation in Eq.\,(\ref{newnew}), the parameter $\mu_{12}^2$ cannot be smaller than $(1\,\rm GeV)^2$, which implies a considerable suppression of the expected gravitational wave signal for $v_\ell \sim 10 \  {\rm PeV}$ in this model, making it very unlikely to detect such a domain wall signature in any near-future experiment.

\section{Conclusions}

Gravitational wave experiments opened an entirely new window of opportunities for probing particle physics models via searches for signatures of first order phase transitions, cosmic strings, and domain walls. 
Already at this point, the sensitivity of the LVK detectors grants access to regions of parameter space  far beyond the reach of any conventional high energy physics experiment. With 
new gravitational wave experiments planned for construction in the near future, sensitive to a much wider range of frequencies, as well as the upcoming improvements to the existing LVK detectors, the search for physics beyond the Standard Model will certainly 
intensify and become even more exciting.

In this work we demonstrated how to exploit the upcoming gravitational wave experiments: Einstein Telescope,  Cosmic Explorer, DECIGO, and Big Bang Observer, to search for signatures of models explaining simultaneously two of the most pressing open questions in particle physics -- the nature of dark matter and the overwhelming domination of matter over antimatter in the present Universe. The solution to the second  puzzle requires a period of an out-of-equilibrium dynamics in the early Universe. This can be realized by a first order phase transition, which is precisely the process whose signatures  gravitational wave detectors are sensitive to. This shows the increasing importance  of gravitational wave experiments for this branch of particle physics in the years to come.

We focused on a representative model of asymmetric dark matter, in which the Standard Model symmetry is extended by a gauged ${\rm SU}(2)_\ell$. In this theory, the baryon number excess is generated through a novel type of instanton interactions. With the  symmetry breaking scale for the new gauge group at $\sim \mathcal{O}(1\!-\!1000) \ {\rm PeV}$, this model does not alter the Standard Model predictions in collider experiments. Nevertheless, as we have shown,  such a high symmetry breaking scale makes it an ideal candidate for gravitational wave searches, with a potential of finding its signatures in all four aforementioned near-future experiments.

A natural continuation of this project would be to consider theories of asymmetric dark matter based on other gauge extensions of the Standard Model, and to develop strategies to differentiate between their gravitational wave signatures. Some examples of such models include a theory based on an ${\rm SU}(4)$ gauge group unifying color and gauged baryon number \cite{Fornal:2015boa}, or a theory based on ${\rm SU}(5)$ where color is unified with a dark ${\rm SU}(2)_{D}$ \cite{Murgui:2021eqf}. One could also investigate other asymmetric dark matter theories with extra  ${\rm U}(1)$ gauge groups \cite{Shelton:2010ta,vonHarling:2012yn}, for which an additional cosmic string contribution would be present in the gravitational wave spectrum.

\subsection*{Acknowledgments} 

We are grateful to the {\emph{Physical Review D}} referee for very constructive comments regarding the manuscript. 
This research was supported by the National Science Foundation\break under Grant No. PHY-2213144.

\bibliography{bibliography}

\end{document}